\documentclass[aps,prl,twocolumn,superscriptaddress,showpacs,preprintnumbers,bibliography]{revtex4-1}
\usepackage[colorlinks,bookmarks=false,citecolor=blue,linkcolor=red,urlcolor=blue]{hyperref}
\usepackage{physics}
\input pdfcolor.tex
\usepackage{colortbl,amsthm,txfonts}
\usepackage{verbatim}
\usepackage{graphicx}
\usepackage{epsfig}
\usepackage{dcolumn}

\usepackage{bm}

\usepackage{amsmath,amssymb}

\makeatletter
\renewcommand*\env@matrix[1][*\c@MaxMatrixCols c]{%

\hskip -\arraycolsep
\let\@ifnextchar\new@ifnextchar
\array{#1}}
\makeatother
\usepackage{appendix}

\begin{document}

\title{Straintronics across Lieb-Kagome interconversion and variable transport scaling exponents} 
\author{Shashikant Singh Kunwar}
\affiliation{Zhejiang Institute of Modern Physics, Zhejiang University, Hangzhou, China}
\affiliation{Department of Physics and Astronomy, University of Iowa, \\
Iowa City, Iowa 52242, United States of America}
\author{Madhuparna Karmakar}
\email{madhuparna.k@gmail.com}
\affiliation{Department of Physics and Nanotechnology, SRM Institute of Science and Technology,   \\ 
Kattankulathur, Chennai 603203, India}

\begin{abstract}
We propose a novel protocol of low-temperature, strain-tuned re-entrant metal-insulator transition and crossover between strongly correlated line-graph lattices (Lieb and Kagome). Using a non-perturbative numerical approach, we demonstrate for the first time 
that an applied shear strain stabilizes a metallic phase cradled in between a gapped magnetic insulator and a gapless flat band localized insulator, facilitating the Lieb/Kagome interconversion. Our results on transport signatures exhibit variable scaling exponents for electrical resistivity and optical conductivity, providing clear evidence of non-Fermi liquid physics. We also define 
a strain-dependent thermal scale to quantify the crossover between the non-Fermi liquid and bad metal phases. 
\end{abstract}

\date{\today}
\maketitle

\textit{Introduction:}
Straintronics, the external strain-controlled engineering of material functionalities, when combined with 
strong electronic correlations, promises exotic quantum phenomena and their novel applications. This synergy
is particularly compelling in the context of two-dimensional (2D) quantum materials with momentum-independent electronic dispersion, such as, Kagome and Lieb materials, which have recently emerged as promising candidates for exploring the physics of many-body correlations \cite{swart_natphys2017,wu_sciadv2018,molina_prl2015,takahashi_sciadv2015,desyatnikov_prl2016,zaanen_nature2015,huang_natcom2020,liu_natcom2019,xu_natphys2018,yazyev_natphys2019,hasan_nature2022}.  Despite the observation of 
novel correlation effects, such as interaction-controlled flat-band-induced electronic localization in Kagome metals 
\cite{checkelsky_natphys2024}, the straintronics of flat band quantum materials remain poorly understood. 
This is largely due to: $(i)$ the lack of tunable material platforms to systematic control  quantum correlation 
effects, and $(ii)$ the absence of suitable numerical framework to simulate the low-temperature physics of such materials. 
This letter focuses on $(ii)$, while providing a prospective protocol for the experimental realization of $(i)$. 
\begin{figure}
\begin{center}
\includegraphics[height=6.5cm,width=8.5cm,angle=0]{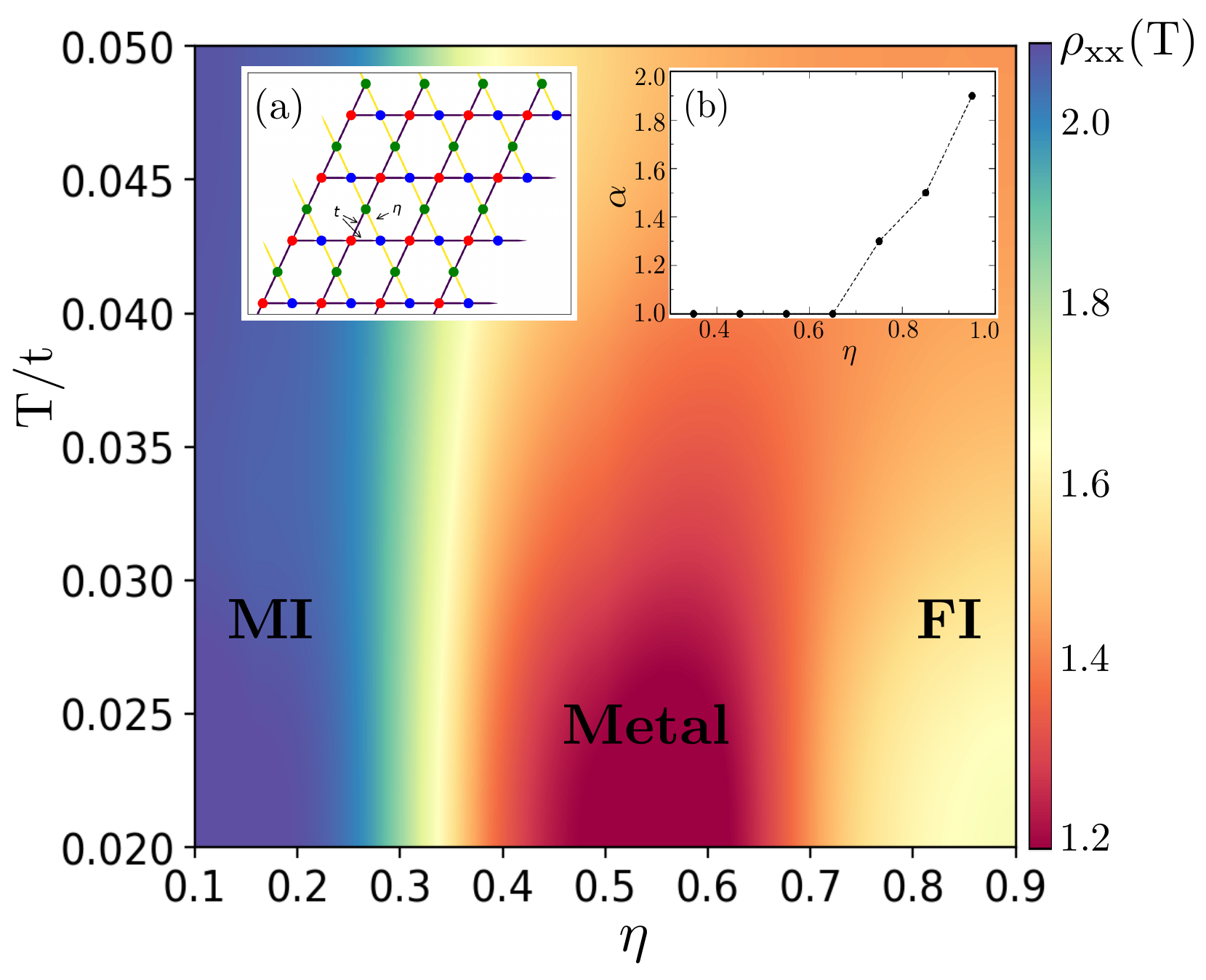}
\caption{Resistivity map in the $\eta-T$ plane demarcating the MI and FI phases at small and large strains, respectively and 
a NFL metal dome at intermediate strain. The color intensity indicates the magnitude of $\rho_{xx}$. The inset: (a) shows the 
lattice structure highlighting the hopping ($t$) and the strain ($\eta$) parameters; the shaded region corresponds to the unit cell, 
(b) shows the strain-dependent variable resistivity scaling exponent across the metal and insulator phases.} 
\label{fig1}
\end{center}
\end{figure}

Amongst the candidate materials, of particular interest are the metal-organic framework (MOF), and covalent organic framework (COF), whose properties are highly tunable via judicious selection of the constituent components and external perturbations \cite{schiffrin_natcom2024,clerac_natcom2022,schiffrin_advfuncmat2021,medhekar_npjcompmat2022,bredas_mathor2022,hasimoto_sciadv2021,heine_chemsocrev2020,feng_chemsocrev2021}. Quantum-correlated 2D MOFs with Kagome or Lieb geometry are currently at the forefront of  material engineering research exhibiting, unconventional superconductivity \cite{hasimoto_sciadv2021}, exchange correlation induced ferromagnetism \cite{medhekar_npjcompmat2022}, and most recently gate-controlled Mott metal-insulator transition \cite{schiffrin_natcom2024}. The host of flat bands, emergent Dirac cones, and non trivial van Hove singularities, the Kagome/Lieb lattices being line graph lattices, are interconvertible \cite{liu_prb2019,denx_pra2023,pereira_prb2023,huang_natcom2020,liu_natcom2019,montambaux_prb2020}. A prospective protocol for the experimental realization of this interconversion is  through strain engineering of 2D MOFs  \cite{low_nanolett2020}. 
 
In this letter,  we for the first time quantify and analyze the straintronics of the shear-strain-tuned interconversion between the Kagome and Lieb lattices in terms of their spectroscopic and transport characteristics  at low temperatures. The system is modeled as the 2D Hubbard Hamiltonian with the nearest neighbor hopping mimicking the shear strain. Quantum correlations are addressed using the non-perturbative,  static path approximated (SPA) Monte Carlo technique (see supplementary materials (SM)), which suitably captures the low-energy physics across the interconversion in the interaction-strain-temperature space  \cite{shashi_kagome2024}. Our main results from this work entails: $(i)$ a strain-tuned {\it re-entrant} metal-insulator transition and crossover, stabilizing a low-temperature non-Fermi liquid (NFL) metal at intermediate strain. $(ii)$ In the large strain regime,  a {\it strain tuned} variable resistivity scaling exponent corresponding to the flat band controlled {\it transient localization} of the itinerant fermions. $(iii)$ Our optical  conductivity results define a strain-dependent Ioffe-Regel-Mott thermal scale ($T_{IRM}(\eta)$) quantifying the crossover to the bad metal phase, characterized by the temperature dependence of the high frequency displaced Drude peak (DDP). 
 
\textit{Model and Method:}
We model the system using the 2D Hubbard Hamiltonian \cite{janson_prb2021,paiva_prb2023,shashi_kagome2024}, 
\begin{eqnarray}
\hat H & = & -\sum_{\langle ij \rangle, \sigma} t_{ij}(\hat c_{i, \sigma}^{\dagger}\hat c_{j, \sigma} + h. c. ) -\eta\sum_{\langle ij \rangle, \sigma} (\hat c_{i, \sigma}^{\dagger}\hat c_{j, \sigma} + h. c.) \nonumber \\ && - \mu\sum_{i, \sigma} \hat n_{i\sigma} + U \sum_{i}\hat n_{i, \uparrow}\hat n_{i,\downarrow} \nonumber \\ && 
\end{eqnarray}

where, $t_{ij} = t=1$ is the hopping integral between the nearest neighbors and sets the reference energy scale of the problem, $\eta$ quantifies the applied shear strain to traverse between the Lieb ($\eta=0$) and the Kagome ($\eta=1$) limits, as shown in the inset of Fig.\ref{fig1}. $U > 0$ is the on-site Hubbard repulsion and is fixed to be $U=t$. We work with the half filled lattice maintained by adjusting the chemical potential $\mu$. The model is made numerically tractable via Hubbard-Stratonovich (HS) decomposition \cite{hs1,hs2} of the interaction term,  introducing a random, fluctuating vector ${\bf m}_{i}(\tau)$ and a scalar $\phi_{i}(\tau)$ (bosonic) auxiliary fields at each site, which couples to the spin and the charge channels, respectively \cite{shashi_kagome2024}. Within the purview of SPA we work in the adiabatic (slow boson) regime, where the bosonic fields can be treated as classical.  The slow bosons serve as a static, randomly fluctuating {\it disordered} background to the fast moving fermions. The approximation is valid for $T \gtrsim \omega_{0}$, where $\omega_{0}$ is the characteristic frequency of the fluctuating bosons \cite{ciuchi_scipost2021,fratini_prb2023,kivelson_pnas2023}. The adiabatic approximation enables us to access the real frequency ($\omega$) dependent observables without requiring an analytic continuation, an edge over the existing non perturbative numerical techniques.  However, the approximation is ill suited for $\omega < \omega_{0}$ and therefore should not be used for $T < T_{FL}$, where $T_{FL}$ is the Fermi liquid (FL) temperature \cite{shashi_kagome2024,ciuchi_scipost2021,fratini_prb2023,kivelson_pnas2023}. 

Within our scheme of calculation the $\phi_{i}$ field is treated at the saddle point level as $\phi_{i} \rightarrow \langle \phi_{i}\rangle = \langle n_{i}\rangle U/2$ (where, $\langle n_{i}\rangle$ is the number density of the fermions), while the complete spatial fluctuations of ${\bf m}_{i}$ are retained. The fermionic correlator characterizing the phases, include: $(i)$ electronic spectral function ($A({\bf k}, \omega)$), $(ii)$ optical conductivity ($\sigma(\omega)$) and $(iii)$ electrical resistivity ($\rho_{xx}$) (see SM). The results presented in this letter corresponds to a system size of $3\times L^{2}$, with $L=18$, and are verified to be robust against finite system size effects.  
\begin{figure}
\begin{center}
\includegraphics[height=5.5cm,width=8.2cm,angle=0]{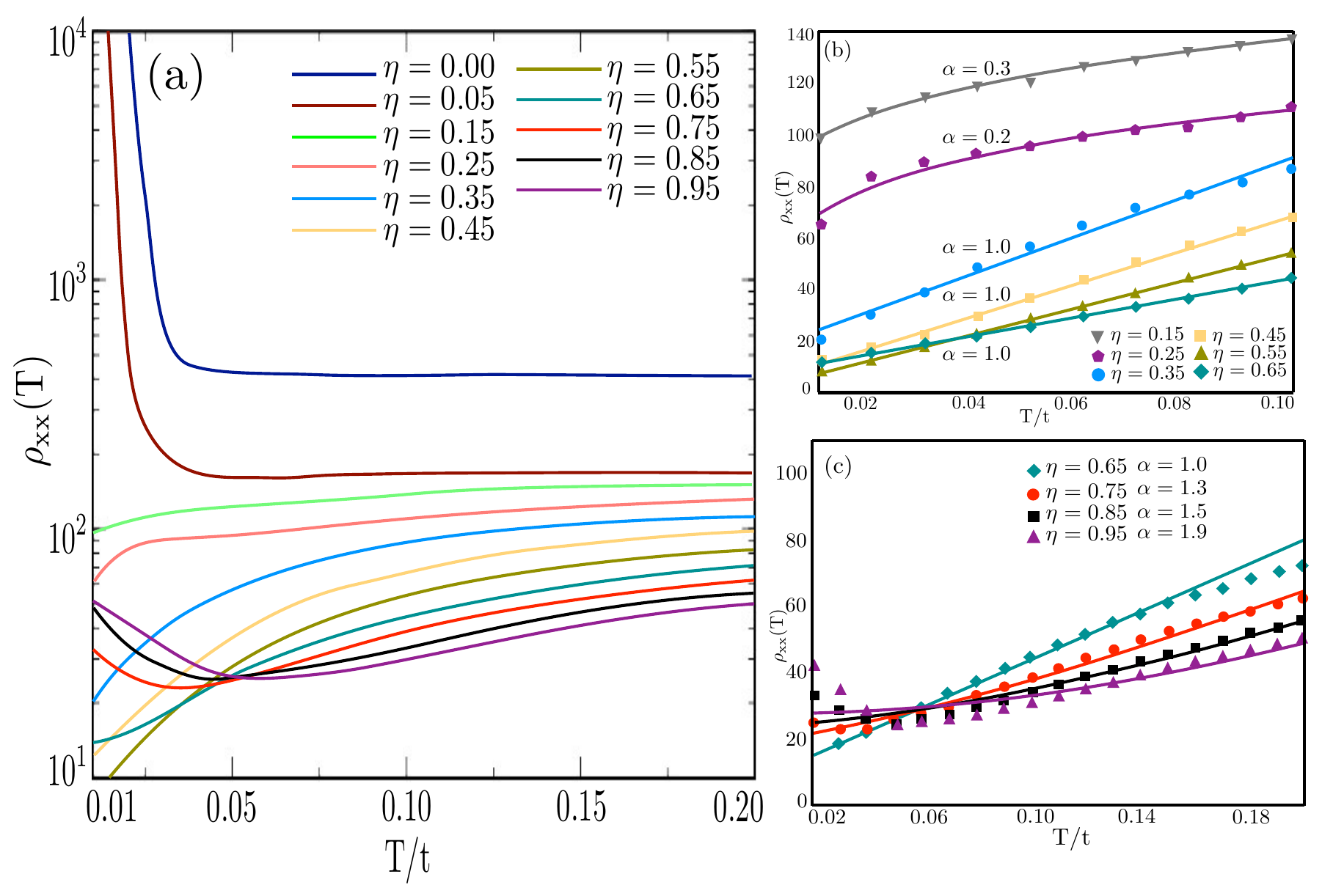}
\caption{(a) Temperature dependence of $\rho_{xx}(T)$ at selected $\eta$. The slope $d\rho_{xx}/dT > 0$ corresponds 
to a metal at intermediate $\eta$ while for the MI and FI regime $d\rho_{xx}/dT < 0$. (b-c) Resistivity (points) in the low 
temperature regime, fitted with $\rho_{xx} = \rho(0)+AT^{\alpha}$ (solid line), where $\alpha(\eta)$ mentioned in the legends 
correspond to the strain dependent resistivity scaling exponent.}
\label{fig2}
\end{center}
\end{figure}

\textit{Re-entrant metal-insulator transition and crossover:}
Fig.\ref{fig1} summarizes the principal result of this work in terms of the resistivity ($\rho_{xx}(T)$) map in the $\eta-T$ plane. The transition and crossover demarcating the low temperature phases are quantified based on Fig.\ref{fig2}(a), which shows the temperature dependence of $\rho_{xx}$ at selected $\eta$ values. In particular, the metallic and insulating phases are distinguished based on the sign of $d\rho_{xx}/dT$, where $d\rho_{xx}/dT > 0$ indicates a metallic phase, while $d\rho_{xx}/dT < 0$ corresponds to an insulating phase. The non-interacting Lieb lattice ($\eta=0$) is a ferromagnetic metal with significant spectral weight at the Fermi level (due to the flat band), interaction gives way to a gapped (ferro) magnetic insulator (MI), with $d\rho_{xx}/dT < 0$ in the small strain regime,  $\eta \lesssim 0.15$. At $\eta=1$,  the non-interacting Kagome lattice consists of a high-energy flat band at $\omega = 2t$ and two Dirac bands that intersect at the ${\bf K}$-point. In the weak coupling regime at low temperatures, the system is magnetically disordered and exhibits a transiently localized insulating phase \cite{shashi_kagome2024}. This localization is a natural outcome of the dynamically disordered, fluctuating local moment (thermal boson) background to which the electrons are subjected \cite{ciuchi_scipost2021,kivelson_pnas2023}. For the Kagome materials, a plausible source of these local moments is the flat band. Interactions systematically shift the flat band towards the Fermi level, allowing the local moments to interact with the itinerant fermions. The latter undergo transient localization, resulting in a prominent suppression in the charge transport, {\it \'{a} la} Anderson localization due to dynamic, inherent disorder potential \cite{checkelsky_natphys2024,shashi_kagome2024}. 
The insulating phase,  i.e., the flat band insulator (FI) arising from such localization, is expected to be gapless, akin to the Anderson mechanism. In Fig.\ref{fig1} the large strain FI regime of $\eta \gtrsim 0.7$ is characterized by $d\rho_{xx}/dT < 0$, as shown in Fig.\ref{fig2}(a). Since the disorder effect is dynamic, it is washed out for $T < T_{FL}$, when the translation invariance of the system is restored, and the conventional FL metal is recovered across $0 \le \eta < 1$. 

Cradled between the two insulating phases, MI and FI, a strain controlled metallic phase is stabilized in the intermediate strain regime, $0.15 \lesssim \eta < 0.7$, quantified by $d\rho_{xx}/dT >0$, giving rise to a re-entrant behavior. A second-order insulator-metal transition (IMT) takes place at $\eta \sim 0.15$, as evidenced by the collapse of the MI gap and the change in the sign of $d\rho_{xx}/dT$. At intermediate strain, metallicity systematically increases with $\eta$, giving rise to a metallic dome with strongly suppressed $\rho_{xx}$ (Fig.\ref{fig1}). At $\eta \sim 0.7$, a re-entrant crossover (sans quantum symmetry breaking) to the FI phase is observed, marked by a change in the sign of $d\rho_{xx}/dT$ (Fig.\ref{fig2}(a)). A detailed investigation of the metal and insulator phases in terms of relevant topological signatures is beyond the scope of this work; however, we do not rule out their possible topological attributes, owing to the emergent Dirac cones \cite{denx_pra2023,pereira_prb2023,montambaux_prb2020}.
\begin{figure}
\begin{center}
\includegraphics[height=7.8cm,width=8.5cm,angle=0]{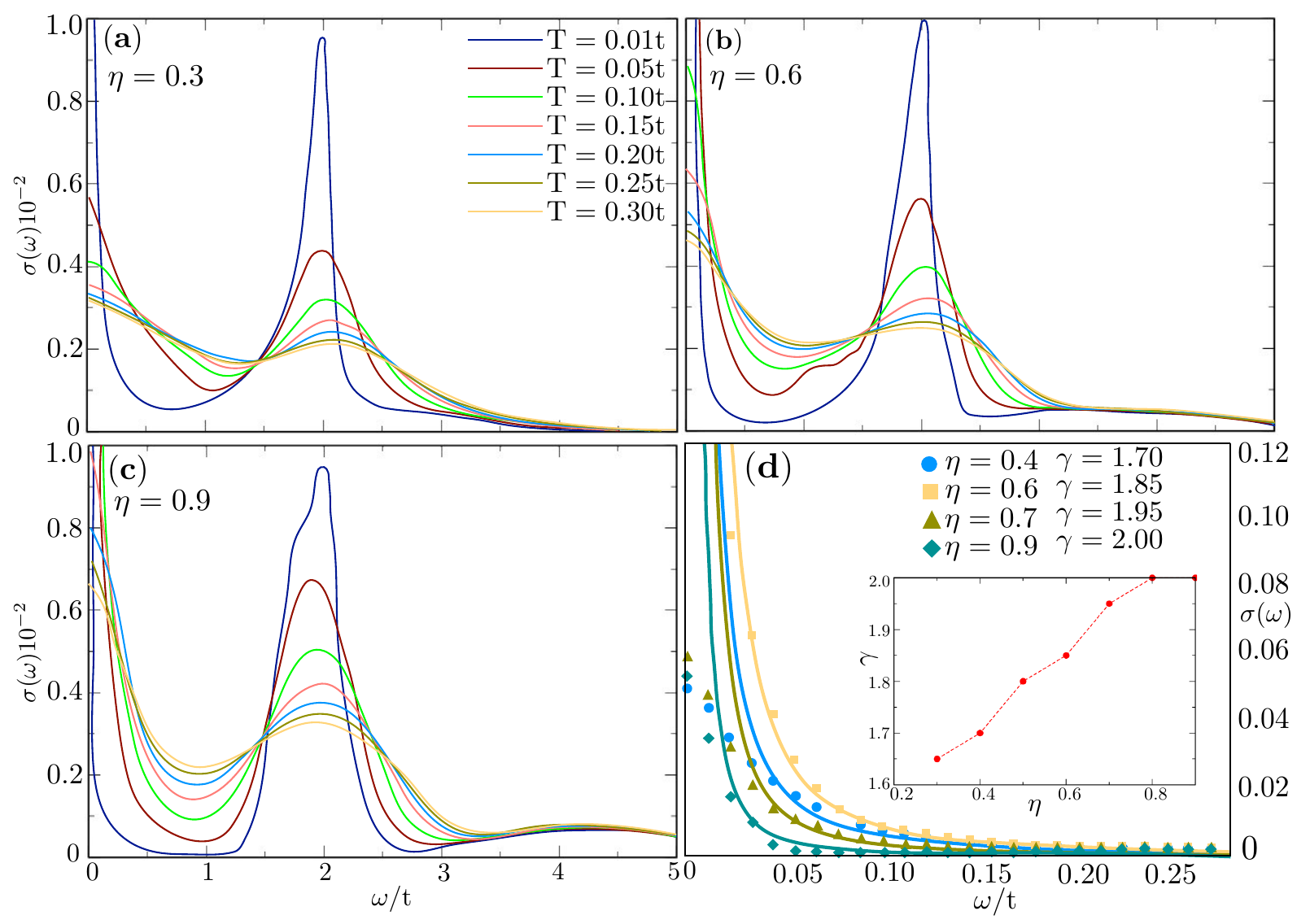}
\caption{(a-c) Temperature dependence of the optical conductivity $\sigma(\omega)$ at selected strains of 
$\eta = 0.3, 0.6$ and $0.9$, comprising of the DDP and high-frequency polaronic peak. (d) The low-frequency DDP 
(points) fitted to $\sigma(\omega) \propto 1/\omega^{\gamma}$ (solid line). The inset shows the strain dependence 
of the variable optical conductivity scaling exponent $\gamma(\eta)$.}
\label{fig3}
\end{center}
\end{figure}

\textit{Resistivity scaling exponent:}
We now analyze the metal and insulating phases across the strain-tuned transition and crossover based on the resistivity scaling exponents. As per Landau's FL theory, conductivity in a metal scales with the quasiparticle lifetime $\tau$, which decays exponentially as the Fermi surface becomes notional. The corresponding resistivity exhibits $\rho_{xx} \propto T^{2}$ behavior in the FL regime, $T < T_{FL}$. In strongly correlated quantum materials, resistivity is often found to violate the $\propto T^{2}$ dependence, suggesting that the underlying state is a NFL \cite{fratini_natcom2021,kanoda_prl2020,mak_nature2021,pasupathy_nature2021,tsirlin_prb2021,pustogow_prb2023,ciuchi_scipost2021}.  Fig.\ref{fig2}(a) demonstrates that the temperature dependence of $\rho_{xx}$ can be classified into two distinct regimes, segregated by a broad shoulder like feature corresponding to a strain-dependent temperature, say $T^{*}(\eta)$, which marks the change in the slope $d\rho_{xx}/dT$. This is in agreement with the existing consensus on the transport signatures of the Hubbard model \cite{devereaux_science2019}. For $T > T^{*}$, $\rho_{xx}$ is roughly independent of temperature and is only weakly dependent on the strain. In contrast, the $T < T^{*}$ regime exhibits explicit dependence of $\rho_{xx}$ on both temperature and strain, which is quantified in terms of the resistivity scaling exponents. We find that $\rho_{xx}$, when fitted with $\rho_{xx} = \rho(0)+AT^{\alpha}$, exhibits a variable sub-linear dependence on $T$ near the IMT, such that for $\eta = 0.15$ and $0.25$; 
$\alpha \sim 0.3$ and $\sim 0.2$, respectively, as shown in Fig.\ref{fig2}(b). Deep in the metallic regime $0.4 \lesssim \eta < 0.7$, a {\it linear-T} dependence (i. e. $\alpha=1$) is observed, attesting that the underlying phase is a NFL metal. 

Fig.\ref{fig2}(c) shows $\rho_{xx}(T)$ in the large strain regime $\eta > 0.7$, fitted to $\rho_{xx} = \rho(0)+AT^{\alpha}$. A variable resistivity scaling exponent is observed in this regime, such that for $\eta=0.75, 0.85, 0.95$; $\alpha \sim 1.3, 1.5, 1.9$, respectively. Experimental observation of variable resistivity scaling exponent between $\alpha \sim 1 - 1.6$ have previously been reported for heavy fermions \cite{assmus_prl1998,steglich_prl2000,lonzarich_jpcm1996,ramazashvili_jpcm2001}. Moreover, Anderson insulators with disorder-localized single-particle states \cite{imry_book2002} and disorder-controlled metal-Mott insulator transitions \cite{dagotto_prl2017} are also known to exhibit a variable resistivity scaling exponent. The origin of the variable resistivity exponent observed in the FI regime of Fig.\ref{fig1} can be traced back to the transient localization of the fermions by the thermal bosons. We observe that the localization is particularly strong near the Kagome limit, leading to insulating behavior, possibly due to the proximity of the flat band to the Fermi level. The strain dependence of the resistivity scaling exponent, $\alpha$ is shown as the inset of Fig.\ref{fig1}. 

\textit{Optical transport and bad metal:}
One of the principal signatures of breakdown of the FL description in strongly correlated materials is the observation of DDP 
in the optical conductivity. While the physical origin of this DDP continues to be debated 
\cite{karlsson_scipost2017,grilli_prl2002,han_rmp2003,takagi_philmag2004,georges_prl2013,georges_prb2013,devereaux_science2019,dobro_prl2015,ciuchi_scipost2021,fratini_prb2023,kivelson_pnas2023}, we argue that the transient localization scenario provides the required insight into this issue. The suppression of the low-frequency charge transport at low temperatures due to dynamic localization results in the shift of the Drude peak to the higher frequencies \cite{ciuchi_scipost2021,ciuchi_arxiv2024}. 
We show the thermal evolution of $\sigma(\omega)$ in Fig.\ref{fig3}(a-c),  at selected $\eta = 0.3, 0.6$, and $0.9$.  For a conventional FL state, the Drude peak decays as $\propto 1/\omega^{2}$ \cite{georges_prl2013,pruschke_prb1995}. Fig.\ref{fig3}(d) shows the low-frequency DDP,  fitted to $\sigma(\omega) \propto 1/\omega^{\gamma}$, where $\gamma$ is a strain-dependent variable optical conductivity scaling exponent. For $\eta=0.3, 0.6$ and $0.9$, $\gamma=1.65, 1.8$ and $2$,  respectively.  The deviation from $\gamma = 2$ behavior is consistent with the underlying NFL phase, in agreement with our observations on 
$\rho_{xx}$ \cite{georges_prl2013,georges_prb2013}. Note that the $\gamma \sim 2$ in the large strain regime corresponds to a possible scaling exponent and does not imply a FL state. 

In addition to the DDP, a high-frequency polaronic peak is observed (Fig.\ref{fig3}(a-c)), arising from the strong correlation between the bosonic background and the itinerant fermions \cite{ciuchi_scipost2021,ciuchi_arxiv2024}. Temperature leads to the overall broadening of the optical spectra and the eventual loss of the DDP via spectral weight transfer. The high-temperature regime comprises of fluctuating local moments with short-range correlations between them, which undergo suppression with temperature. The corresponding optical spectrum is rather featureless, except for a broad high-frequency hump. The loss of the DDP indicates the crossover to the bad metallic regime and defines the scale, $T_{IRM}(\eta)$, which quantifies the thermal crossover scale between the NFL and the bad metal phases \cite{han_rmp2003,takagi_philmag2004}. While, for the intermediate strain regime, the crossover to the bad metal takes place from a NFL metal, in the large and small strain regimes, $T_{IRM}(\eta)$ marks this crossover from NFL insulating phases. For $\eta=0.3$, $T_{IRM}(\eta) \sim 0.2t$,  while for $\eta =0.6$, $T_{IRM}(\eta) \sim 0.25t$. At $\eta=0.9$, the NFL regime survives over the entire range of temperature probed in this work, suggesting that $T_{IRM}(\eta) \propto \eta$.  Similar crossover physics were observed in 2D transition metal dichalcogenides (TMD) \cite{trivedi_prl2014,mak_nature2021,pasupathy_nature2021}, organic charge transfer salts \cite{fratini_natcom2021,kanoda_prl2020}, rare earth nickelates \cite{stemmer_sciadv2021} and Kagome metals  \cite{tsirlin_prb2021}, undergoing disorder, doping, temperature, or gate controlled metal-insulator transitions. Our observations highlight applied strain as a potential control parameter for the NFL-bad metal crossover physics. 
\begin{figure}
\begin{center}
\includegraphics[height=8.3cm,width=8.3cm,angle=0]{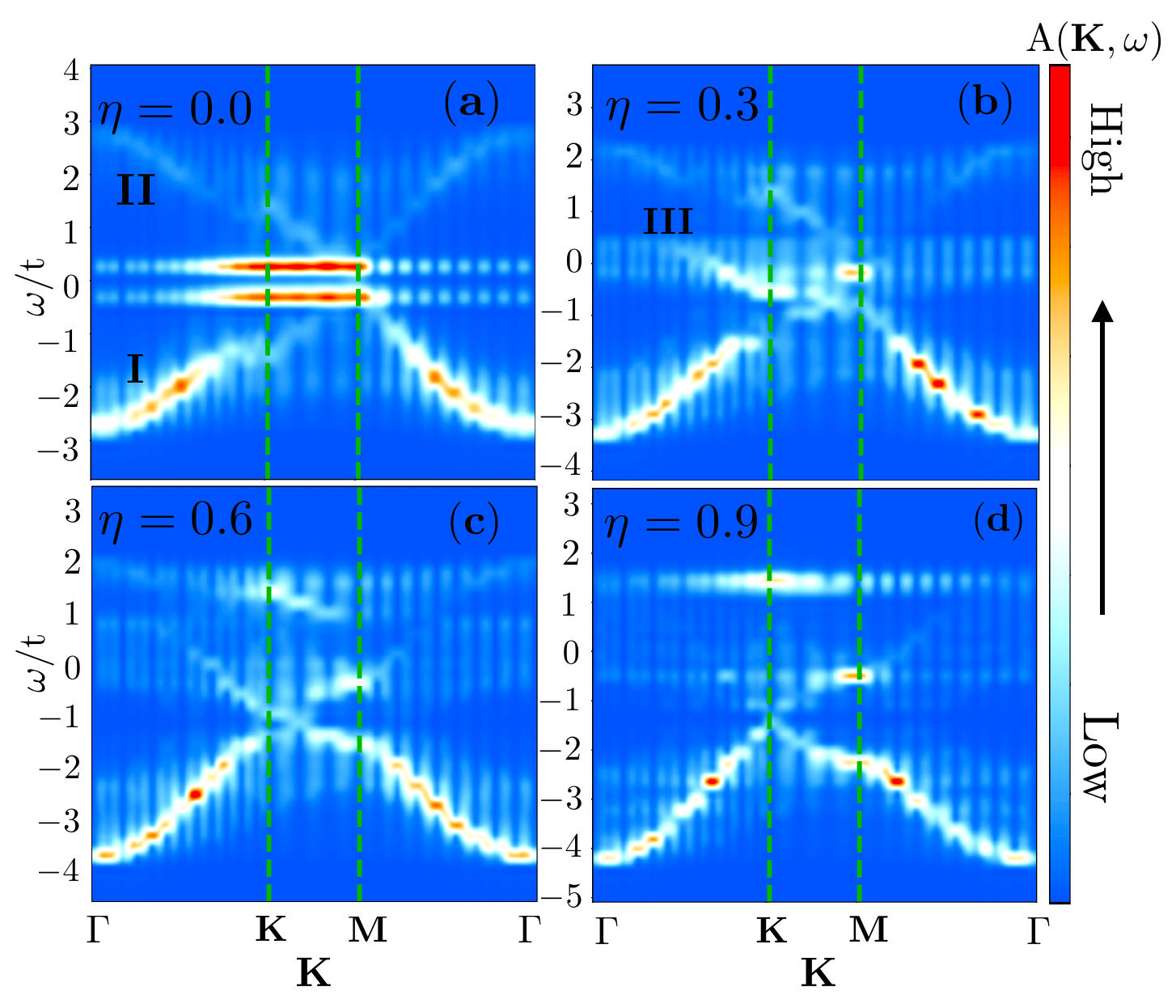}
\caption{The spectral function $A({\bf k}, \omega)$ at $T=0.01t$ for $\eta=0.0, 0.3, 0.6$ and $0.9$, representative of the 
small, intermediate and large strain regimes. Notable, we observe a strain-controlled reconstruction of the electronic band 
structure and the emergence of Dirac cones across the lattice interconversion.}
\label{fig4}
\end{center}
\end{figure}

\textit{Strain controlled band structure engineering:}
The prospect of strain-tuned transitions and crossover of the electronic states is particularly appealing because, unlike chemical substitution, strain does not introduce randomness into the system. Instead, it  can modify or preserve the lattice symmetry in a controlled manner. We discuss the corresponding band structure engineering in light of the electronic spectral function $A({\bf k}, \omega)$, shown in Fig.\ref{fig4} at selected values of $\eta$. For convenience, we designate the lower and upper Dirac bands as bands I and II, respectively. At the $\eta=0$ limit (Fig.\ref{fig4}(a)), the system is characterized by an insulating gap at the Fermi 
level. This gap collapses with small strain (Fig.\ref{fig4}(b)), giving way to a gapless phase. The flat band now becomes momentum-dependent, and the Dirac point begins to shift from the $M$-point towards the $K$-point. As the lattice is strained towards the Kagome limit, a new Dirac band (say, band-III) emerges via the splitting of the reminiscent flat band at the Fermi level, as shown in Fig. \ref{fig4}(b). This new band intersects band-I between the $M$ and the $K$-points. The residual flat band at the Fermi level move to higher energies, convoluting with band-II to form a new band with weak momentum dependence.  

At intermediate strain, represented by $\eta=0.6$ in Fig.\ref{fig4}(c), three processes take place simultaneously: $(i)$ The Dirac point between the bands I and III shifts from the $M$ to the $K$-point, $(ii)$ the overall spectrum shifts to lower energies, and $(iii)$ the emergent high-energy band becomes momentum-independent. The spectrum in the large strain regime of $\eta > 0.7$ is akin to what is expected from the Kagome geometry with a high energy flat band and two Dirac bands intersecting each other at the $K$-point (Fig.\ref{fig4}(d)).  While the band structure reconstruction and the corresponding Dirac point shifts have been probed in significant detail in the non interacting limit across the Lieb/Kagome interconversion \cite{denx_pra2023,pereira_prb2023,montambaux_prb2020}, our results for the first time illustrate how these shifts give rise to novel quantum phases and exotic phase transitions or crossovers in the presence of electronic correlations. 

\textit{Discussion and conclusion:}
The results of this work are obtained using SPA Monte Carlo technique, based on the approximation of classical (slow) bosons. In this framework, fermions are subjected to a static, randomly fluctuating, disordered, bosonic background. We argue that a finite (low) temperature transiently localized regime is an expected consequence of itinerant fermions interacting with the thermal bosons, providing an explanation for the breakdown of the FL description. The stability of this regime is dictated by the characteristic boson frequency of the system, $\omega_{0}$,  while the strength of localization depends on the material parameters such as electronic interaction, applied strain, lattice geometry, and more. In terms of the transport signatures, we infer that while a NFL phase emerges across strain-applied regime, it can be further classified into insulating and metallic phases based on the strength of localization. 
SPA is ill-suited for capturing the $T=0$ physics of the system, particularly in the gapless regime. However, there should always be 
a finite temperature transiently localized regime above a coherence temperature $T_{FL}$ in strongly correlated materials, as discussed herein. SPA, therefore provides important insight on the low-temperature physics of the system, which remains inaccessible to most of the existing numerical techniques due to severe fermionic sign problem and strong finite system size effects. 

The primary impact of neglecting of quantum fluctuations in this scheme is the absence of the FL regime in the transport signatures. Our resistivity results show linear-$T$ dependence and a variable resistivity scaling exponent in the low-temperature metallic and gapless insulating regimes, respectively. At $T \rightarrow 0$, a $\rho_{xx} \propto T^{2}$ behavior is expected for $T < T_{FL}$, corresponding to the FL metal, as experimentally observed in several classes of strongly correlated quantum materials \cite{fratini_natcom2021,kanoda_prl2020,mak_nature2021,pasupathy_nature2021,tsirlin_prb2021}. Similarly, the low-frequency Drude peak in the optical conductivity with $\propto 1/\omega^{2}$ dependence can not be captured within the purview of SPA. 
In the absence of any calculation which takes into account the ``quantumness'' of the bosonic fields in geometrically frustrated lattices at low temperatures, a direct comparison with our results can't be made. However, while such a calculation would accurately capture the $T < T_{FL}$ regime, we do not expect any qualitative changes in the results discussed here for $T > T_{FL}$.

One of the key results of this work is the strain-stabilized NFL metal that undergoes re-entrant metal-insulator transition and crossover. Our observation of strain dependent Anderson-like localization and the corresponding deviation from the $\rho_{xx} \propto T$ behavior close to the Kagome limit aligns well with the results obtained from the recent transport measurements on Kagome metal,  Ni$_{3}$In \cite{checkelsky_natphys2024,si_natcom2024}. The origin of the sub-linear behavior of the resistivity exponent in the small strain regime, close to the IMT, is more subtle. Its possible origin could be topological in nature, arising out of the emergent Dirac cones across the strain tuned band structure reconstruction (see SM) \cite{denx_pra2023,pereira_prb2023,montambaux_prb2020}. Another important result from this work is the strain-controlled finite temperature crossover between the NFL and the bad metal phases, quantified in terms of a strain-dependent thermal crossover 
scale $T_{IRM}(\eta)$.  Such thermal crossovers have been reported in the context of correlated materials undergoing bandwidth, disorder and gate-controlled IMT  \cite{fratini_natcom2021,kanoda_prl2020,trivedi_prl2014,mak_nature2021,pasupathy_nature2021,stemmer_sciadv2021,tsirlin_prb2021}. In this letter,  we  introduce ``applied strain'' as a potential tuning parameter for the quantum phases. Our straintronic protocol is generic and should be applicable to other natural and engineered strongly correlated quantum materials, particularly those with line-graph lattices.   

Among the possible experimental avenues, MOF and COF holds  particular appeal due to their broad tunability in terms of composition, structural transition and choice of substrates, leading to their advance functionalities \cite{cui_accchemres2016}. In 2D MOF, strain can be introduced either via lattice mismatch with the substrate or by mechanically straining the MOF-substrate composite \cite{medhkar_npjcompmat2022,goronzy_acsnano2018,guinea_jpcm2015}. The Lieb and Kagome lattices are the two fundamental structural building blocks of 2D MOFs, and they have been experimentally realized in materials such as, 9, 10-dicyanoanthracene-copper (DCA-Cu) \cite{yan_advfuncmat2021,schiffrin_advfuncmat2021,yan_acsnano2021}, MCl$_{2}$(pyrazine)$_{2}$ \cite{clerac_natcom2022}, 1, 2, 4, 5-tetracyanobenzene (TCNB) \cite{schneider_acsnano2024} etc. Charge transfer, gate tuning, strain engineering etc. are shown to bring forth the effects of strong electronic correlations in these materials \cite{yan_advfuncmat2021,zhou_rscadv2014,dreher_acsnano2021,dendzik_prb2017,guinea_jpcm2015,schiffrin_natcom2024}. 
The results presented in this letter provide, for the first time, the impetus to strain-tune strongly correlated quantum phases across lattice interconversion in 2D MOFs.

\textit{Acknowledgements:} MK would like to acknowledge  the use of the high performance computing facility (AQUA) at the Indian 
Institute of Technology, Madras, India. MK acknowledges the support from the Department of Science and Technology, Govt. of India 
through the grant CRG/2023/002593.

\section{Supplementary Material}

\textit{Model, method and indicators:}
Our starting Hamiltonian is the repulsive Hubbard model on a Kagome lattice, defined as, 
\begin{eqnarray}
\hat H & = & -\sum_{\langle ij \rangle, \sigma} t_{ij}(\hat c_{i, \sigma}^{\dagger}\hat c_{j, \sigma} + h. c. ) -\eta\sum_{\langle ij \rangle, \sigma} (\hat c_{i, \sigma}^{\dagger}\hat c_{j, \sigma} + h. c.) \nonumber \\ && - \mu\sum_{i, \sigma} \hat n_{i\sigma} + U \sum_{i}\hat n_{i, \uparrow}\hat n_{i,\downarrow} \nonumber \\ && 
\end{eqnarray}

\noindent here, $t_{ij} = t=1$ is the hopping integral between the nearest neighbors and sets the reference energy scale of the problem. $\eta$ quantifies the applied shear strain that transitions between the Lieb ($\eta=0$) 
and  Kagome ($\eta=1$) limits. $U > 0$ is the on-site Hubbard repulsion and is fixed at $U=t$. We consider the half-
filled lattice, which is maintained by adjusting the chemical potential $\mu$.  To make the model numerically tractable, 
we decompose the interaction term using Hubbard-Stratonovich (HS) transformation  \cite{hs1,hs2}.  
This introduces two (bosonic) auxiliary fields viz. a vector field ${\bf m}_{i}(\tau)$ and a scalar field 
$\phi_{i}(\tau)$, which couple to the spin and charge densities, respectively. The introduction of these auxiliary 
fields preserves the spin rotation invariance, the Goldstone modes, and enables us to capture the Hartree-Fock theory 
at the saddle point. In terms of the Grassmann fields $\psi_{i\sigma}(\tau)$, we write,

\begin{eqnarray}
\exp[U\sum_{i}\bar\psi_{i\uparrow}\psi_{i\uparrow}\bar\psi_{i\downarrow}\psi_{i\downarrow}] & = & \int {\bf \Pi}_{i}
\frac{d\phi_{i}d{\bf m}_{i}}{4\pi^{2}U}{\exp}[\frac{\phi_{i}^{2}}{U}+i\phi_{i}\rho_{i}+\frac{m_{i}^{2}}{U} 
\nonumber \\ && -2{\bf m}_{i}.{\bf s}_{i}]
\end{eqnarray}

where, the charge and spin densities are defined as $\rho_{i} = \sum_{\sigma}\bar\psi_{i\sigma}\psi_{i\sigma}$ 
and ${\bf s}_{i}=(1/2)\sum_{a,b}\bar \psi_{ia}{\bf \sigma}_{ab}\psi_{ib}$, respectively. The corresponding 
partition function then takes the form,

\begin{eqnarray}
{\cal Z} & = & \int {\bf \Pi}_{i}\frac{d\bar\psi_{i\sigma}d\psi_{i\sigma}d\phi_{i}d{\bf m}_{i}}{4\pi^{2}U}
\exp[-\int_{0}^{\beta}{\cal L}(\tau)d\tau]
\end{eqnarray}
where, the Lagrangian ${\cal L}$ is defined as,
\begin{eqnarray}
{\cal L}(\tau) & = & \sum_{i\sigma}\bar\psi_{i\sigma}(\tau)\partial_{\tau}\psi_{i\sigma}(\tau) + H_{0}(\tau) 
\nonumber \\ && +\sum_{i}[\frac{\phi_{i}(\tau)^{2}}{U}+(i\phi_{i}(\tau)-\mu)\rho_{i}(\tau)+
\frac{m_{i}(\tau)^{2}}{U} \nonumber \\ && -2{\bf m}_{i}(\tau).{\bf s}_{i}(\tau)]
\end{eqnarray}
where, $H_{0}(\tau)$ is the kinetic energy contribution. 
The $\psi$ integral is now quadratic, but this comes at the cost of an additional integration over
the fields ${\bf m}_{i}(\tau)$ and $\phi_{i}(\tau)$. The weight factor for the ${\bf m}_{i}$
and $\phi_{i}$ configurations can be determined by integrating out the $\psi$ and
$\bar \psi$ fields. Using these weighted configurations, one can then go back and computes
the fermionic properties. Formally,

{\begin{eqnarray}
{\cal Z} & = & \int {\cal D}{\bf m}{\cal D}{\phi}e^{-S_{eff}\{{\bf m},\phi\}}
\end{eqnarray}}
\begin{eqnarray}
S_{eff}\{{\bf m}, \phi\} & = & \log Det[{\cal G}^{-1}\{{\bf m},\phi\}] + \frac{\phi_{i}^{2}}{U} +
\frac{m_{i}^{2}}{U}
\end{eqnarray}
where, ${\cal G}$ is the electron Green's function in a $\{{\bf m}_{i},\phi_{i}\}$ background.

The weight factor for an arbitrary space-time configuration $\{{\bf m}_{i}(\tau), \phi_{i}(\tau)\}$ 
involves computing the fermionic determinant in that background. If we express the auxiliary fields in terms 
of their Matsubara modes ${\bf m}_{i}(\Omega_{n})$ and $\phi_{i}(\Omega_{n})$, the various approximations 
can then be recognized and compared.

\begin{itemize}
\item{Determinant Quantum Monte Carlo (DQMC): DQMC retains the full ``$i, \Omega_{n}$'' dependence of ${\bf m}$ and $\phi$. 
The approach is valid at all temperatures but does not readily yield real-frequency ($\omega$) properties. Furthermore, for many quantum 
materials, the low-temperature regime is inaccessible to DQMC due to the fermionic sign problem, which is particularly 
severe for magnetically frustrated systems, such as, Kagome lattice. Additionally, the computational cost of DQMC 
restricts its use to small system sizes, leading to strong finite-size effects in the results.}
\item{Homogeneous Mean Field Theory: This theory is time-independent, completely neglects the fluctuations, and replaces the auxiliary 
fields by their mean values to minimize the free energy. Specifically, ${\bf m}_{i}(\Omega_{n}) \rightarrow \vert m\vert$ and 
$\phi_{i}(\Omega_{n}) \rightarrow \vert \phi \vert$. The inhomogeneous Hartree-Fock mean field theory accounts 
for spatial fluctuations in the amplitude of ${\bf m}_{i}$ and $\phi_{i}$, but neglects the angular fluctuations, i. e. ${\bf m}_{i}(\Omega_{n}) \rightarrow \vert m_{i}\vert$ and $\phi_{i}(\Omega_{n}) \rightarrow \phi_{i}$. For $T\neq$0 this approximation breaks down beyond the weak coupling regime.}
\item{Static Path Approximation (SPA): SPA retains the full spatial dependence in ${\bf m}$ and $\phi$ but only keeps the
$\Omega_{n}=0$ mode, i. e. ${\bf m}_{i}(\Omega_{n}) \rightarrow {\bf m}_{i}$ and $\phi_{i}(\Omega_{n}) \rightarrow \phi_{i}$.
It thus includes classical fluctuations of arbitrary magnitudes but no quantum ($\Omega_{n} \neq 0$) fluctuations. 
The different temperature regimes are characterized as, (a) $T=0$: Since classical fluctuations vanish at $T=0$, SPA reduces to standard 
Hartree-Fock mean field theory. (b) At $T \neq 0$: We consider not just the saddle point configuration, but {\it all configurations}
following the weight $e^{-S_{eff}}$ discussed earlier. This approach suppresses the order much faster than in mean field theory. (c) High $T$: Since the $\Omega_{n}=0$ mode dominates the exact partition function, the SPA approach becomes exact as $T \rightarrow \infty$.}
\item{For completeness we mention that the standard dynamical mean field theory (DMFT) retains the full dynamics but 
keeps ${\bf m}$ and $\phi$ at effectively one site, i.e.,  ${\bf m}_{i}(\Omega_{n}) \rightarrow {\bf m}(\Omega_{n})$ and $\phi_{i}(\Omega_{n}) \rightarrow \phi(\Omega_{n})$. This is exact in the limit $D \rightarrow \infty$, where $D$ is the spatial dimension.}
\end{itemize}

Following the SPA approach, we freeze $\phi_{i}(\tau)$ to its saddle point value 
$\phi_{i}(\tau)=\langle n_{i} \rangle U/2$, where $\langle n_{i}\rangle$ is the fermionic 
number density. The resulting model can be viewed  as fermions coupled to spatially fluctuating random background of classical field ${\bf m}_{i}$. With 
these approximations, the effective Hamiltonian corresponds to a coupled spin-fermion model and reads as, 

\begin{eqnarray}
H_{eff} & = & -t\sum_{\langle ij\rangle, \sigma}[c_{i\sigma}^{\dagger}c_{j\sigma}+h.c.] +\eta\sum_{\langle \langle ij \rangle \rangle}
[\hat c_{i, \sigma}^{\dagger}\hat c_{j, \sigma} + h. c.]\nonumber \\ && -\tilde \mu \sum_{i\sigma} \hat n_{i\sigma}  - \frac{U}{2}\sum_{i}{\bf m}_{i}.{\bf \sigma}_{i} + \frac{U}{4}\sum_{i}m_{i}^{2}
\end{eqnarray}
where, $\tilde \mu =\sum_{i}(\mu-\langle n_{i}\rangle U/2)$, and the last term of $H_{eff}$ corresponds to the 
stiffness cost associated with the classical field 
${\bf m}_{i}$. Here, ${\bf \sigma}_{i}=\sum_{a,b}c_{ia}^{\dagger}{\bf \sigma}_{ab}c_{ib}={\bf s}_{i}$.

The random background configurations of $\{{\bf m}_{i}\}$ are generated numerically via Monte Carlo 
simulations and follow the Boltzmann distribution,
\begin{eqnarray}
P\{{\bf m}_{i}\} \propto Tr_{c,c^{\dagger}}e^{-\beta H_{eff}}
\end{eqnarray}

For large and random configurations, the trace is computed numerically by diagonalizing $H_{eff}$ for each attempted 
update of ${\bf m}_{i}$, and the system converges to the equilibrium configuration using Metropolis algorithm. Evidently, this process is computationally 
expensive. 

\textit{Indicators:} Once the equilibrium configurations of $\{{\bf m}_{i}\}$ are obtained, the different phases are characterized based 
on the following fermionic correlation functions, 

\begin{itemize}
\item{Optical conductivity, calculated using the Kubo formula, 
\begin{eqnarray}
\sigma(\omega) & = & \frac{\sigma_{0}}{N} \sum_{\alpha, \beta} \frac{f(\epsilon_{\alpha})-f(\epsilon_{\beta})}{\epsilon_{\beta}-\epsilon_{\alpha}} \vert \langle \alpha \vert J_{x} \vert \beta \rangle \vert^{2} \delta(\omega - (\epsilon_{\beta}-\epsilon_{\alpha})) \nonumber \\ 
\end{eqnarray}
where, the current operator $J_{x}$ is defined as, 
\begin{eqnarray}
J_{x} & = & -i\sum_{i, \sigma, \vec \delta} [{\vec \delta}t_{\vec \delta}c_{{\bf r}_{i}, \sigma}^{\dagger}c_{{\bf r}_{i}+\vec \delta, \sigma} - H. c.]
\end{eqnarray}
The dc conductivity ($\sigma_{dc}$) is the $\omega \rightarrow 0$ limit of $\sigma(\omega)$, with $\sigma_{0}=\frac{\pi e^{2}}{\hbar}$ in 2D. $f(\epsilon_{\alpha})$ is the Fermi function, and $\epsilon_{\alpha}$ and $\vert \alpha\rangle$ are the single-particle eigenvalues and eigenvectors of $H_{eff}$, respectively, in a given background of $\{{\bf m}_{i}\}$.
}
\item{Spectral function, 
\begin{eqnarray}
A({\bf k}, \omega) & = & -(1/\pi){\mathrm{Im}}G({\bf k}, \omega)
\end{eqnarray}
where, $G({\bf k}, \omega) = lim_{\delta \rightarrow 0} G({\bf k}, i\omega_{n})\vert_{i\omega_{n} \rightarrow \omega + i\delta}$.  $G({\bf k}, i\omega_{n})$ is the imaginary frequency transform of $\langle c_{\bf k}(\tau)c_{\bf k}(0)^{\dagger}\rangle$.  
}
\end{itemize}

\textit{Benchmarking SPA:}
The static path approximation has been  extensively used to investigate various quantum many-body phenomena, such as, 
the BCS-BEC crossover in superconductors \cite{tarat_epjb}, Fulde-Ferrell-Larkin-Ovchinnikov (FFLO) superconductivity in solid 
state systems and ultracold atomic gases \cite{mpk_imb,mpk_mass}, the Mott transition in frustrated lattices 
\cite{rajarshi,nyayabanta_pyrochlore,nyayabanta_chk,nyayabanta_epl,mpk_spinliq}, d-wave superconductivity 
\cite{dagotto_prl2005}, competition and coexistence of magnetic and d-wave superconducting orders \cite{dagotto_prb2005}, 
orbital-selective magnetism relevant for iron superconductors \cite{dagotto_prb2016}, strain-induced superconductor-insulator transition in flat band lattices \cite{lieb_strain}, among others. In many of these problems, the use of numerically exact techniques like 
DQMC is not feasible, either due to the sign problem or to the system size limitations (particularly for multiband systems). Thus, judicial approximations are essential, and SPA is one such method that can capture the low-temperature phases and the thermal properties of these strongly correlated systems with reasonable accuracy. 
\begin{figure*}
\begin{center}
\includegraphics[height=4.2cm,width=13.5cm,angle=0]{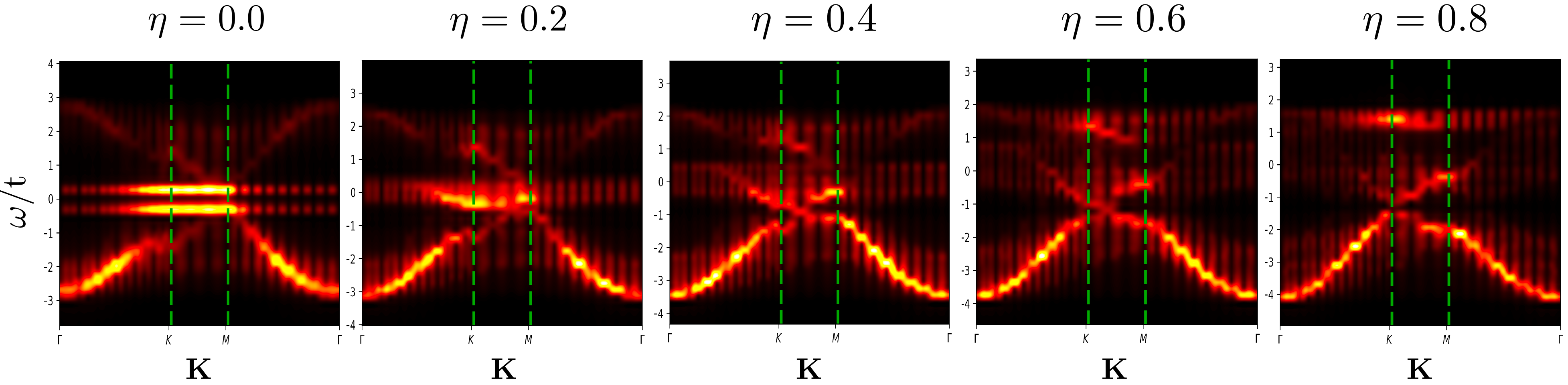}
\caption{Strain tuned reconstruction of the electronic spectra across the Lieb/Kagome interconversion.}
\label{fig1_suppl}
\end{center}
\end{figure*}

Strong correlation combined with geometric frustration present a formidable challenge to most existing non-perturbative numerical 
techniques, particularly at low temperatures. The large-strain regime of our model corresponds to the Kagome lattice, a highly frustrated 2D 
structure featuring flat band and non-trivial van Hove singularities. This construction significantly limits the choice of the available techniques 
to address the physics of these systems. There are primarily two categories of calculations reported in the literature. The first focusses exclusively on the ground state properties of the system using specialized numerical approaches, such as DMRG \cite{zhu_prbl2021} and functional renormalization group (FRG) \cite{thomale_prr2024}. The second explores the high-temperature phases of the Kagome metal and 
insulator using sophisticated non-perturbative numerical approaches like DQMC, DMFT and its variants\cite{janson_prb2021}. However, due to 
severe fermionic sign problem and system size limitations, these techniques cannot be pushed to sufficiently low temperatures. Moreover, 
approaches like single-site DMFT doesn't take into account the short range spatial fluctuations of the bosonic fields, which, as shown in the main text, are crucial for capturing localization physics and consequent breakdown of the FL theory. Quantitative comparisons 
have shown that DMFT, in comparison to the adiabatic boson approximation, fails to capture the low-temperature physics 
of generic strongly correlated quantum materials with sufficient accuracy \cite{ciuchi_scipost2021}. Extrapolating high-temperature and 
$T=0$ results to address the low-temperature regime is therefore not feasible and can potentially lead to incorrect inferences 
\cite{shashi_kagome2024}. On the other hand, SPA provides an important tool for addressing this problem and offers crucial insights into the 
low-temperature physics of the correlated quantum materials. It also allows for easy accessibility of the real-frequency dynamical observables for sufficiently
large system sizes, offering a significant edge over existing techniques.    

\textit{Strain tuned band structure reconstruction:}
We present a strain-controlled systematic reconstruction of the electronic band structure across the Lieb/Kagome interconversion at 
$U=t$ and $T=0.01t$, shown in Fig.\ref{fig1_suppl}. In the Lieb limit ($\eta=0$), the system behaves as a gapped magnetic insulator, with the interaction splitting the flat band at the Fermi level and creating two Dirac bands that touch at the  $M$-point. In the small-strain regime, the flat band distorts and accumulates spectral weight at the Fermi level, rendering the spectra gapless. The band touching point shifts away from the $M$-point. At intermediate strain, a high-energy flat band emerges, and  the band touching point evolves into a band crossing at the $K$-point. The Fermi level now contains significant spectral weight, while the overall spectra shifts to lower energies. As the system approaches the Kagome limit ($\eta=1$) at large strain, the electronic spectra consist of a high-energy flat band that moves closer to the Fermi level compared to its non-interacting counterpart. The two Dirac bands crosses at the $K$-point. The applied strain not only shifts the crossing of the Dirac bands from the $M$ point to the $K$-point, but also leads to the deconstruction of the flat band at the Fermi level and its subsequent reconstruction at a higher energies, across the Lieb/Kagome interconversion. 

\bibliography{reentrant_arxiv.bib}
\end{document}